\documentclass{article}
\usepackage[dvips]{graphicx}
\usepackage{spconf,amssymb,amsmath,epsfig,color,url}
\usepackage{algorithm,algorithmic}
\usepackage{listings}

\newcommand{\beq}{\begin{equation}}
\newcommand{\eeq}{\end{equation}}
\newcommand{\beqn}{\begin{eqnarray}}
\newcommand{\eeqn}{\end{eqnarray}}

\def\bmath#1{\mbox{\boldmath$#1$}}

\floatname{algorithm}{Algorithm}

\long\def\symbolfootnote[#1]#2{\begingroup%
\def\thefootnote{\fnsymbol{footnote}}\footnote[#1]{#2}\endgroup}

\title{A Stochastic LBFGS Algorithm for Radio Interferometric Calibration}

\name{Sarod Yatawatta$^{\star}$ \thanks{This work was supported by Netherlands eScience Center (project DIRAC, grant 27016G05).} \qquad Lukas De Clercq$^{\dagger}$ \qquad Hanno Spreeuw$^{\dagger}$ \qquad Faruk Diblen$^{\dagger}$}

\address{$^{\star}$ ASTRON, The Netherlands Institute for Radio Astronomy,\\ Dwingeloo, The Netherlands. Email: yatawatta@astron.nl\\ 
    $^{\dagger}$Netherlands eScience Center, Science Park 140 (Matrix I),\\ 1098 XG Amsterdam, The Netherlands.
}
\begin{document}
\ninept
\maketitle
\begin{abstract}
We present a stochastic, limited-memory Broyden Fletcher Goldfarb Shanno (LBFGS) algorithm that is suitable for handling very large amounts of data. A direct application of this algorithm is radio interferometric calibration of raw data at fine time and frequency resolution. Almost all existing radio interferometric calibration algorithms assume that it is possible to fit the dataset being calibrated into memory. Therefore, the raw data is averaged in time and frequency to reduce its size by many orders of magnitude before calibration is performed. However, this averaging is detrimental for the detection of some signals of interest that have narrow bandwidth and time duration such as fast radio bursts (FRBs). Using the proposed algorithm, it is possible to calibrate data at such a fine resolution that they cannot be entirely loaded into memory, thus preserving such signals. As an additional demonstration, we use the proposed algorithm for training deep neural networks and compare the performance against the mainstream first order optimization algorithms that are used in deep learning.
\end{abstract}
\section{Introduction}
Modern radio interferometers produce raw data at a very fine time and frequency resolution covering a large observing bandwidth and a long observing time. This means that the total number of data points at the original resolution can run into billions, thus entering radio astronomy into the big-data era. In order to apply most calibration algorithms (e.g., \cite{Boonstra03,Jeffs06,Wijn,Kaz2,Kaz3,Kazemi3,ICASSP13,Tasse,DCAL,DMUX,SIRP,Brossard2018}) the number of data points needs to be reduced in size to fit into memory. This is done by averaging the data in frequency and in time, thus reducing the resolution of the data. While this makes the data manageable, it also has some drawbacks. First, signals of scientific interest with a narrow bandwidth and time duration such as FRBs \cite{Chat2017} might be lost. Second, the removal of low-power, broad-band interfering signals (both terrestrial and celestial origin) \cite{Fridman,aoflagger,Soko2016} might be difficult.

In this paper, we address the calibration of radio interferometric data at the original resolution that is determined by the correlator. Calibration is essential and preferable at this resolution for the removal of strong interfering signals such as the Sun and a few other strong radio sources (e.g., Cassiopeia A, Cygnus A). Naturally, the amount of data being calibrated will not fit into memory and in order to solve the nonlinear optimization problem associated with calibration, we introduce a stochastic, limited-memory Broyden Fletcher Goldfarb Shanno (LBFGS) algorithm. Due to faster convergence than gradient descent methods and lower memory usage than methods that require the full Hessian \cite{Fletcher,Liu1989}, LBFGS has been extensively used in our calibration software \cite{InPar}. 

In LBFGS, the direction of descent is estimated by using the gradient of the cost function and an approximation to the (inverse) Hessian of the cost function, represented by a few norm $1$ updates \cite{Fletcher}. Once the descent direction is found, the amount of descent is determined by selecting the step size based on a line search algorithm. The approximation to the Hessian is updated with each iteration, by using the difference in the parameters and the difference in the gradients (curvature pairs) \cite{Fletcher,NW}. There are several problems that need to be overcome in adopting LBFGS to multi-batch mode operation.
First, there is increased noise in gradient estimation using a mini-batch compared to the use of the full dataset. Secondly, and more critically, calculating the difference of gradients cannot in principle use two different mini-batches for gradient estimation. We reiterate that our main constraint is the size of the dataset and we aim to devise an algorithm that can load fixed size mini-batches into memory. The novelty of the proposed method and the relation to prior work are as follows.
\begin{itemize}
\item Estimating the variance of the gradient is a crucial element both in \cite{Bolla} and our method. This ideally requires the gradient information for each data point within a single mini-batch. However, in our method, we use an online estimate of the variance of the gradient (based on \cite{Welford}) and this does not require the gradient of each data point within a mini-batch. This is the major novelty in our work.
\item In order to find the difference in gradients, data overlap is used in \cite{Schr2007,Berahas}. Our method does not use  data overlap (nor is it practically feasible). However, we do use more than one iteration per mini-batch, which can be considered as full data overlap.
\item Variable mini-batch sizes are used in \cite{Bolla}, however we keep the mini-batch size fixed. 
\item A fixed step size is used in \cite{Schr2007,Berahas} and \cite{Li2018} uses normalized descent directions, both slowing the convergence. As in \cite{Bolla}, our method uses a variable step size based on Armijo line search \cite{NW}. Since we use an online estimate of the variance of the gradient, the initial value of step size for the line search is determined by this value, similar to \cite{Bolla}.
\item We use stable curvature pair updates as in \cite{Berahas,Bolla}. In fact, whenever a new mini-batch is used, we skip the update of the curvature because the difference in gradients is based on different data. This is a drawback of our algorithm that requires more than one iteration per each mini-batch.
\item Our previous work used ordered subsets acceleration in \cite{OS} where we used mini-batches for the gradient (Jacobian) estimation but the cost function was evaluated using the full batch. However, this method does not work when the full batch cannot fit in memory (which is the problem we are concerned with in this work).
\end{itemize}
 
The rest of the paper is organized as follows. We give a brief introduction to radio interferometric calibration using the LBFGS algorithm in section \ref{sec:calib}.  We describe the proposed stochastic LBFGS algorithm in section \ref{sec:alg}. Next, we provide test results based on radio interferometric calibration and deep learning in \ref{sec:simul} illustrating the performance of the proposed algorithm. Finally, we draw our conclusions in section \ref{sec:conclusions}.
Notation: Matrices and vectors are denoted by bold upper and lower case letters as ${\bf J}$ and ${\bf v}$, respectively. The transpose and the  Hermitian transpose are given by $(\cdot)^T$ and $(\cdot)^H$, respectively. The matrix  Frobenius norm is given by $\|\cdot \|$ and the $l_1$ norm by $\|\cdot \|_1$. The set of real and complex numbers are denoted by  ${\mathbb R}$ and ${\mathbb C}$, respectively. The identity matrix (size $N\times N$) is given by ${\bf I}_N$. The Hadamard (element wise) product is denoted by $\circ$.

\section{Radio interferometric calibration}\label{sec:calib}
In this section we give a brief overview of radio interferometric calibration in relation to the use of the LBFGS algorithm. A single data point produced by a pair of stations $p$ and $q$ forming an interferometer can be modeled as \cite{HBS}
\beq \label{V}
{\bf V}_{pq}=\sum_{i=1}^K {\bf J}_{pi} {\bf C}_{pqi} {\bf J}_{qi}^H + {\bf N}_{pq}
\eeq
where ${\bf V}_{pq}$,${\bf J}_{pi}$,${\bf J}_{qi}$,${\bf C}_{pqi}$ and ${\bf N}_{pq}$ are in ${\mathbb C}^{2\times 2}$. All matrices in (\ref{V}) are functions of time and frequency and this is implicitly assumed throughout the rest of the paper. This observed signal ${\bf V}_{pq}$ is modeled as a superposition of electromagnetic radiation emanating from $K$ distinct celestial sources in the sky. In (\ref{V}), ${\bf C}_{pqi}$ represents the source coherency for the $i$-th source, seen from the interferometer (or baseline) $p$-$q$. The values of ${\bf C}_{pqi}$ can be exactly calculated for any given $p,q,i$ and at any time and frequency \cite{TMS}. The signals from celestial sources are corrupted by the atmosphere as well as the instrument and  these corruptions are represented by the matrices given by  ${\bf J}_{pi}$,${\bf J}_{qi}$ in (\ref{V}). We also have the noise matrix ${\bf N}_{pq}$. Calibration is the estimation of ${\bf J}_{pi}$ for all $p$ and $i$, or estimating a set of parameters ${\bmath \theta}$ describing ${\bf J}_{pi}$.

Vectorizing (\ref{V}) we get
\beq \label{vecV}
{\bf v}_{pq}={\bf s}_{pq}({\bmath \theta}) +{\bf n}_{pq}
\eeq
where ${\bf s}_{pq}({\bmath \theta})=\sum_{i=1}^{K}({\bf J}_{qi}^{\star}\otimes{\bf J}_{pi}) vec({\bf C}_{pqi})$ and  ${\bf v}_{pq}=vec({\bf V}_{pq})$, ${\bf n}_{pq}=vec({\bf N}_{pq})$. We represent ${\bf J}_{pi}$ and ${\bf J}_{qi}$ in (\ref{vecV}), as ${\bmath \theta}$, a vector of real parameters of length $8NK$ ($\in {\mathbb R}^{8NK\times 1}$), but this could have more parameters if we add time or frequency dependence to ${\bf J}_{pi}$ and ${\bf J}_{qi}$. We stack data points as vectors
\beqn
{\bf x}=[\mathrm{real}({\bf v}_{12}^T),\mathrm{imag}({\bf v}_{12}^T),\mathrm{real}({\bf v}_{13}^T),\ldots]^T\\\nonumber
{\bf m}({\bmath \theta})=[\mathrm{real}({\bf s}_{12}^T),\mathrm{imag}({\bf s}_{12}^T),\mathrm{real}({\bf s}_{13}^T),\ldots]^T\\\nonumber
\eeqn
where (assuming data at $T$ time and $F$ frequency samples are calibrated together) ${\bf x}$ and  ${\bf m}({\bmath \theta})$ are vectors of size $N(N-1)/2\times 8\times T \times F$ ($\in  {\mathbb R}^{4TFN(N-1)}$).

The cost function minimized by LBFGS in robust calibration is given by 
\beq \label{fLBFGS}
f({\bmath \theta})=\sum_{i=1}^{4TFN(N-1)} \log\left(1 +\frac{\left({\bf x}[i]-{\bf m}({\bmath \theta})[i]\right)^2}{\nu}\right)
\eeq
where ${\bf x}[i]$ and ${\bf m}({\bmath \theta})[i]$ represent the $i$-th elements of ${\bf x}$ and ${\bf m}({\bmath \theta})$, respectively, and $\nu$ $(=2)$ is a constant \cite{Kaz3}. More detail about the use of full batch mode LBFGS in calibration can be found in \cite{InPar}. In the next section, we consider modifications to the original LBFGS algorithm when the full dataset cannot be loaded into memory.

\section{Stochastic LBFGS algorithm}\label{sec:alg}
We describe the modifications made to the batch mode LBFGS algorithm for operation in mini-batch (stochastic) mode in this section. The key driver for these changes is the need to calibrate a dataset that cannot entirely fit into memory, i.e., the length of ${\bf x}$ $4TFN(N-1)$ is too large to fit into memory. The number of parameters or length of ${\bmath \theta}$ $8NK$ is large but not too large that it cannot fit into memory. We use $M$ as the memory size of the LBFGS algorithm, and for the stochastic mode of operation, we need $2$ additional vectors of the size of ${\bmath \theta}$. Thus, in total, the extra memory requirement is $2(M+1)$ vectors of the size of ${\bmath \theta}$.

The pseudocode for the stochastic LBFGS algorithm is given in algorithm \ref{algLBFGS}. We describe in detail key operations that are different from the full batch LBFGS algorithm as follows.
Note that we keep the mini-batch size fixed, and we do not have access to gradient statistics within a mini-batch in our algorithm. The price to pay for this approach is that more that one iteration is needed per each mini-batch (ideally about $4$). In all our practical tests, we keep the memory size for Hessian approximation as $M=7$.
\begin{itemize}
\item Line 4: The mini-batches can be processed in any arbitrary order, even though it is presented as sequential processing.
\item Lines 9-10: Online estimation of variance of gradient is based on \cite{Welford}. The update of ${\bf g}_k$ and ${\bf m}_k$ are only done if a new batch of data is received (line 8). Based on the estimated variance, the upper bound for step size $\overline{\alpha}$ is found (line 11).
\item Line 13: BFGS update using $M$ curvature pairs is done as in \cite{NW}.
\item Lines 15-17: Step size selection is done using Armijo line search as in \cite{Bolla,NW}. In contrast, in the full batch LBFGS algorithm \cite{InPar,escience2018}, we use exact line search with cubic interpolation \cite{Fletcher}.
\item Lines 19, 22: We update curvature pairs  $({\bf y},{\bf s})$ only if the gradient difference on line 21 is based on the same batch and if the inner product over norm ratio is larger than a threshold $\epsilon$. This is similar to the safe update rule used in \cite{Berahas,Bolla}.
\end{itemize}
\begin{algorithm}
\caption{Stochastic LBFGS}
\label{algLBFGS}
\begin{algorithmic}[1]
\REQUIRE Data ${\bf x}=\{{\bf x}_1,\ldots,{\bf x}_B\}$, partitioned into $B$ batches of equal size.
Cost $f_i({\bmath \theta})$ and gradient $\nabla f_i({\bmath \theta})$ for each batch $i\in[1,B]$. 
Initial value of parameters ${\bmath \theta}_0$ (number of parameters is $P$), memory for Hessian approximation $M > 0$, iterations per batch $J$, number of epochs $E$, constants $c_1(=10^{-4})$, $\alpha_0(=1)$, $\epsilon(=10^{-9})$.
\STATE $k \gets 0$,${\bmath \theta} \gets {\bmath \theta}_0$
\STATE ${\bf g}_k \gets {\bf 0}$,${\bf m}_k \gets {\bf 0}$ ($\in \mathbb{R}^{P\times 1}$)
\FOR {$n=$ $1$ to $E$} 
{\FOR {$i=$ $1$ to $B$}
\STATE calculate $\nabla f_i({\bmath \theta})$
{\FOR {$j=1$ to $J$ and $\|\nabla f_i({\bmath \theta})\|$ is finite}
\STATE {batch\_changed} $\gets  (k > 1\ \  \mathrm{and}\ \ j=1)$
\IF {batch\_changed} 
\STATE ${\bf g}_k \gets {\bf g}_{k-1} + \frac{\nabla f_i({\bmath \theta}) -{\bf g}_{k-1}}{k}$
\STATE ${\bf m}_k \gets {\bf m}_{k-1} + (\nabla f_i({\bmath \theta}) -{\bf g}_{k})\circ(\nabla f_i({\bmath \theta}) -{\bf g}_{k-1})$
\STATE $\overline{\alpha}\gets \alpha_0\left({1+\frac{1}{\|\nabla f_i({\bmath \theta})\|}\frac{\|{\bf m}_k\|_1}{k-1}}\right)^{-1}$
\ENDIF
\STATE ${\bf p}_{k} \gets -{\bf H}_k\nabla f_i({\bmath \theta})$ by BFGS update
\STATE $\alpha_k \gets \overline{\alpha}$
\WHILE {$f_i({\bmath \theta}+\alpha_k {\bf p}_{k}) > f_i({\bmath \theta})+\alpha_k c_1 \nabla f_i({\bmath \theta})^T {\bf p}_{k}$} 
\STATE $\alpha_k \gets \alpha_k/2$
\ENDWHILE
\STATE ${\bmath \theta}_{new} \gets {\bmath \theta} + \alpha_k {\bf p}_k$
\IF {not batch\_changed}
\STATE ${\bf s}_k \gets {\bmath \theta}_{new}-{\bmath \theta}$
\STATE ${\bf y}_k \gets \nabla f_i({\bmath \theta}_{new}) -\nabla f_i({\bmath \theta})$
\IF {${\bf y}_k^T {\bf s}_k > \epsilon \|{\bf s}_k\|^2$}
\IF {$k\ge M$}
\STATE discard oldest pair of $({\bf y},{\bf s})$
\ENDIF
\STATE save $({\bf y}_k,{\bf s}_k)$ pair
\ENDIF
\ENDIF
\STATE ${\bmath \theta}\gets {\bmath \theta}_{new}$
\STATE ${k \gets k + 1}$
\ENDFOR
}
\ENDFOR
}
\ENDFOR
\RETURN ${\bmath \theta}$
\end{algorithmic}
\end{algorithm}

In the next section, we test the performance of algorithm \ref{algLBFGS} against current contenders. 

\section{Test results}\label{sec:simul}
We demonstrate the performance of the proposed stochastic LBFGS algorithm in its intended use, i.e., radio interferometric calibration as well as a completely different application, i.e., deep learning in this section.
\subsection{Radio interferometric calibration}
We simulate an array with $N=1024$ stations calibrating data along $K=4$ directions in the sky. We use $T=10$ time samples and a single frequency ($F=1$). Therefore the amount of data points being calibrated is $4N(N-1)T \approx 41.9$ million. The number of parameters estimated is $8NK=32768$, which is much smaller than the number of data points. In real life, the number of data points being calibrated could run upto a few billion and the simulation presented here is only a scaled down version. The systematic errors ${\bf J}_{pi},{\bf J}_{qi}$ in (\ref{V}) are drawn from a complex uniform distribution in $[0,1]$ as $\mathcal{U}(0,1)+\jmath\mathcal{U}(0,1)$. Noise is finally added to the data with a signal to noise ratio $E\{\frac{\|{\bf V}_{pq}\|^2}{\|{\bf N}_{pq}\|^2}\}=40$ and noise is drawn from a complex circular Gaussian distribution.  The $K=4$ point sources have intensities ranging from $1$ to $5$ intensity units and are randomly positioned in the sky. The solutions are initialized to ${\bf J}_{pi}={\bf I}_2$ for all $p$ and $i$.
\begin{figure}[htbp]
\begin{minipage}[b]{0.98\linewidth}
\centering
\centerline{\epsfig{figure=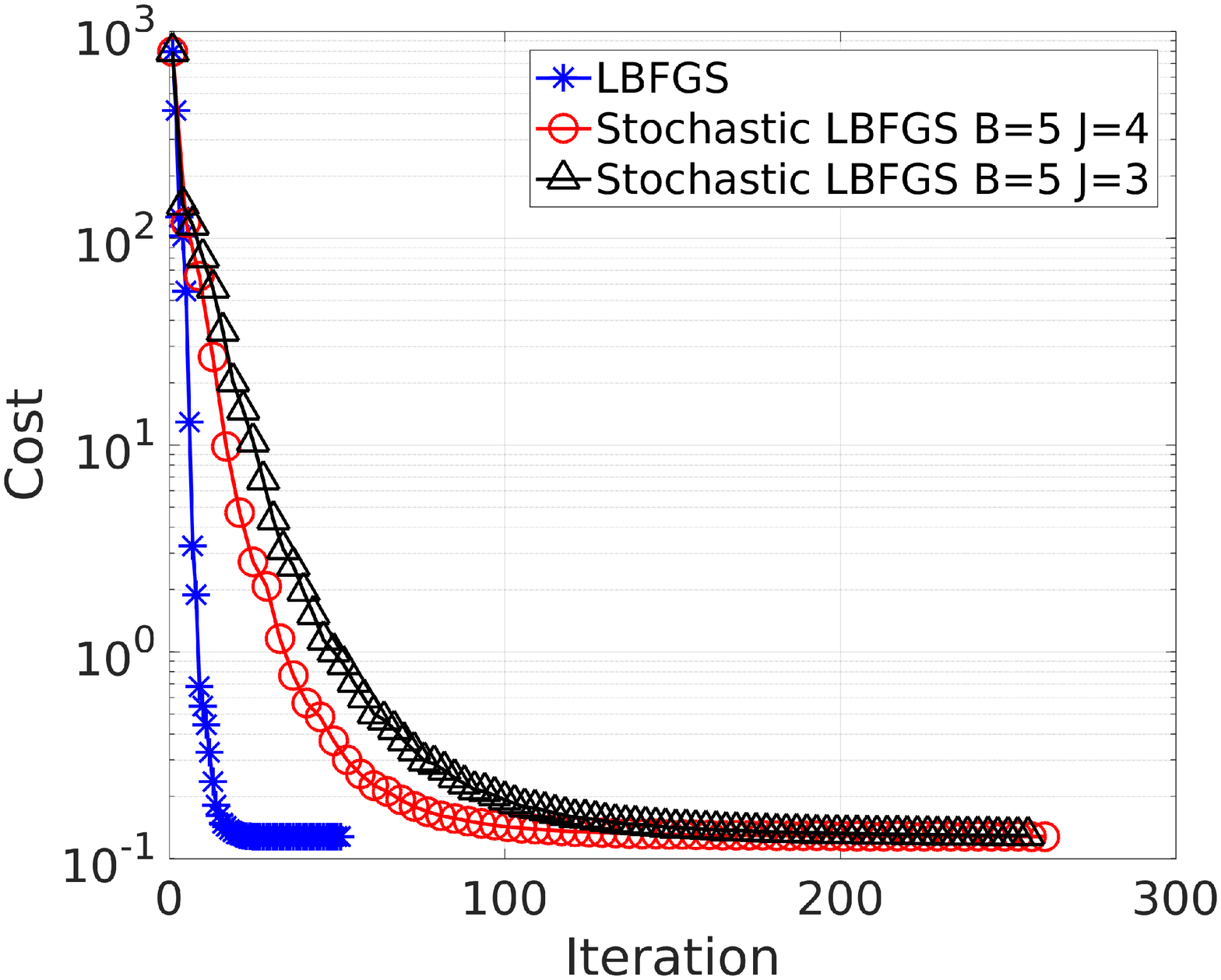,width=7.5cm}}
\end{minipage}
\caption{Calibration error comparison of full batch LBFGS algorithm against stochastic LBFGS algorithm. All algorithms approximately take the same amount of total run time.} \label{fig_cal}
\end{figure}

We compare the performance of LBFGS in full batch mode operation against the stochastic version in calibrating the data  in Fig. \ref{fig_cal}. In full batch mode, we load the whole dataset into memory and use exact line search with cubic interpolation \cite{Fletcher,escience2018} for step size selection. In mini-batch mode operation we use $1/5$-th of the full dataset at each iteration ($B=5$), and for each mini-batch of data, we use $J=3$ ($E=17$) and $J=4$ ($E=13$) LBFGS iterations in algorithm \ref{algLBFGS}. We have shown the cost  (\ref{fLBFGS}) evaluated over the full dataset in Fig. \ref{fig_cal} for comparison. Note that in mini-batch LBFGS, the cost evaluated at each iteration of algorithm \ref{algLBFGS} is actually based on a smaller dataset and what is shown in Fig. \ref{fig_cal} is the cost based on the full dataset, only for comparison.  

From Fig. \ref{fig_cal} we see that we reach the same final cost by using the stochastic version of LBFGS, but the convergence is slower, i.e., many more iterations are needed. Moreover, increasing $J$ (iterations per mini-batch) makes the convergence faster. In terms of computational cost, all three modes of operation in Fig. \ref{fig_cal} are almost equal. For the stochastic version, the cost of computing the cost function and the gradient is $1/5$-th of the cost of computing them with the full dataset. However, more iterations are used in the stochastic mode, thus making the total computational cost almost the same. Nonetheless, the key advantage is that the stochastic mode is using much less data at each iteration.
Note also that we have already compared the use of first order methods in \cite{escience2018} in full batch mode of operation against the full batch mode LBFGS and we see significantly better performance in LBFGS. Therefore we have not presented any such comparisons for radio interferometric calibration here.

\subsection{Deep learning}
The use of the stochastic LBFGS algorithm has so far given mixed results in deep learning \cite{Schr2007,Berahas,Bolla,Li2018} compared to the popular first order methods used in training, i.e., stochastic gradient descent (SGD) \cite{robbins1951} and Adam \cite{Adam}. We have implemented algorithm \ref{algLBFGS} in PyTorch \cite{paszke}, a popular machine learning package. We compare the performance of the proposed algorithm in training a simple convolutional neural network with $2$ convolutional layers and $3$ fully connected layers as given in \cite{TorchCIFAR}. We use the CIFAR 10 dataset \cite{cifar10} and use $50 000$ images for training and $10 000$ images for verification. 

Compared with previous examples of using LBFGS for training \cite{Schr2007,Berahas,Bolla,Li2018}, we have made two changes to the setup. First, we have replaced the Rectified linear unit (ReLU) activation with exponential linear unit (ELU) \cite{ELU} activation to get non-zero second order derivatives. Our second modification is related to improving the computational cost. Automatic differentiation is used in PyTorch for calculating the gradient during training. However, this gradient is not needed every time the cost function is calculated in algorithm \ref{algLBFGS}. Note that the cost function needs to be calculated many times for the line search (lines 15-17 in algorithm \ref{algLBFGS}). This makes the computational cost of LBFGS much larger than SGD or Adam. However, we can improve this by disabling gradient calculation during the line search operation. In order to do this, we have to modify the definition of the loss function ({\tt closure()}) as shown in Fig. \ref{list_closure}.

\begin{figure}[htbp]
\centering
\begin{lstlisting}[language=Python]
for input, target in dataset:
    def closure():
        if torch.is_grad_enabled():
          optimizer.zero_grad()
        output = model(input)
        loss = loss_fn(output, target)
        if loss.requires_grad:
          loss.backward()
        return loss
    optimizer.step(closure)
\end{lstlisting}
\caption{Disabling gradient calculation during line search routines in the definition of the loss function.}\label{list_closure}
\end{figure}

\begin{figure}[htbp]
\begin{minipage}[b]{0.98\linewidth}
\begin{minipage}{0.98\linewidth}
\centering
\centerline{\epsfig{figure=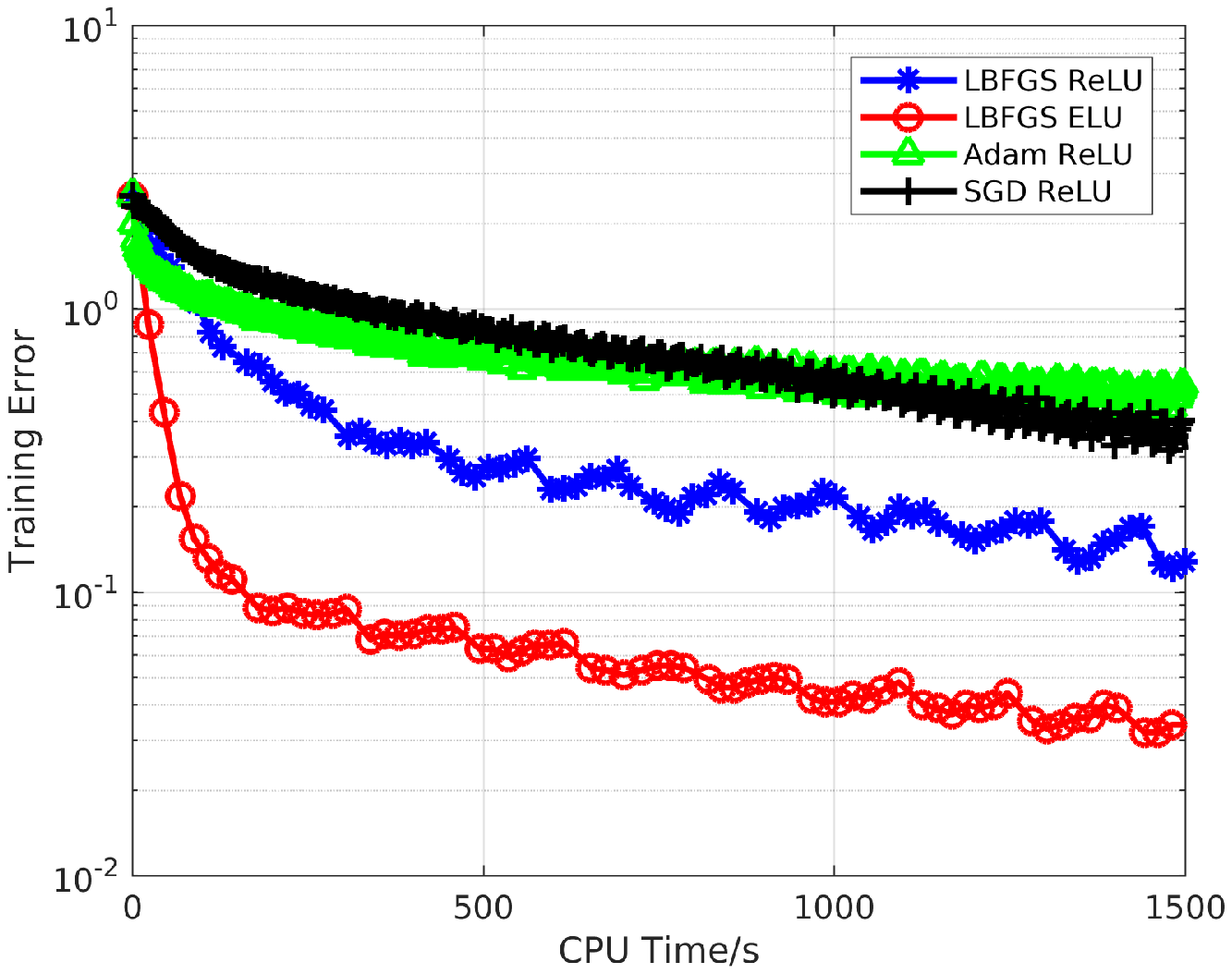,width=7.5cm}}
\end{minipage}
\end{minipage}
\caption{CIFAR 10 training error for mini-batch size 32.} \label{fig_cifar}
\end{figure}

In Fig. \ref{fig_cifar}, we show the training error for the proposed stochastic LBFGS ($M=10$), with ReLU and ELU activation, compared with SGD and Adam (with ReLU activation). We see similar performance with both ReLU and ELU activation for SGD and Adam and we show only one result. The mini-batch size used is $32$ and the training error is shown for every $200$ batches. Per mini-batch, we use $J=4$ iterations in algorithm \ref{algLBFGS}.   

Note that the training error is shown against CPU time used in training. Both SGD and Adam runs about $\times 5$ faster and therefore can process $\times 5$ more epochs than LBFGS for the same amount of total CPU time. The verification accuracy for $10 000$ images is 63\% for LBFGS with ELU activation as well as for SGD and Adam but LBFGS with ReLU activation has a lower verification accuracy of 61\%.  We clearly see the improvement of LBFGS due to ELU activation in Fig. \ref{fig_cifar} and shows the potential of LBFGS for further applications in deep learning.

\section{Conclusions}\label{sec:conclusions}
We have proposed a stochastic LBFGS algorithm that can handle the calibration of very large volumes of radio interferometric data while using a limited amount of compute memory. We have demonstrated the performance of the proposed algorithm by comparing it with the full batch LBFGS algorithm. In addition, we have tested its use in deep learning, where we get comparable performance provided that activation functions are modified. We have implemented the proposed algorithm in PyTorch \cite{paszke} and it is available for general use in other machine learning applications \cite{code_url}. Future work will focus on adopting the proposed algorithm for big-data frameworks as in \cite{Chen2014} for large scale distributed calibration of radio interferometric data especially for the detection of FRBs.

\bibliographystyle{IEEEbib}
\bibliography{references}

\end{document}